# First principles study of edge carboxylated graphene quantum dots

**Hazem Abdelsalam[1], Hanan Elhaes[2], Medhat A. Ibrahim[3]**

[1]Department of Theoretical Physics, National Research Centre, Giza, 12622, Egypt

[2]Physics Department, Faculty of Women for Arts, Science, and Education, Ain Shams University, 11757 Cairo, Egypt

[3]Spectroscopy Department, National Research Centre, Giza, 12622, Egypt

**Abstract:**

The structure stability and electronic properties of edge carboxylated hexagonal and triangular graphene quantum dots are investigated by using density functional theory. The calculated binding energies show that the hexagonal clusters with armchair edges have the highest stability among all other flakes. The binding energy of carboxylated graphene quantum dots increases by increasing the number of attached carboxyl groups. Our study shows that the total dipole moment significantly increases by adding COOH with the highest values observed in triangular clusters. The edge states in triangular graphene with zigzag edges produce completely different energy spectrum from other shapes as (a) the energy gap in triangular zigzag cluster is very small compared to other clusters and (b) the highest occupied molecular orbital is localized at the edges which is in contrast to other clusters where it is distributed over the cluster surface. The enhanced reactivity and the controllable energy gap by shape and edge termination make graphene quantum dots ideal nanodevices for various applications such as sensors. The infrared spectra for different flakes are presented for confirmation and detection of the obtained results.

# 1- INTRODUCTION:

Flake-like graphene quantum dots (GQDs) have attracted great attention due to their unique electronic [1-5] and optical [6-10] properties. The distinguishable properties of GQDs arise from the electron confinement in the finite size graphene cluster that leads to the opening of energy gap and quantization of electronic energy levels. Cutting graphene sheets into small clusters result in creation of GQDs with different shapes and edges. Energy gap strongly depends on shape, edges, and size of the GQDs. Moreover, new states (called edge states) appear in the low energy region that also depend on the size, shape (e.g. hexagonal and triangular), and edge termination (zigzag versus armchair). Tight binding (TB) calculations [1, 2, 5] show that there are two types of edge states in flakes with zigzag termination, zero energy states (ZES) that are degenerated and located exactly at the Fermi level and dispersed energy states (DES) that fill the low energy and are symmetrically distributed around it [1, 2, 5, 11, 12]. ZES appear in triangle GQDs while DES appear in other shapes such as hexagonal and circular GQDs. Due to the inclusion of electron-electron interactions in the density functional theory (DFT) calculations, a tiny energy gap is opened between ZES [2, 6, 9]. In general, TB or DFT calculations confirmed the appearance of edge states in GQDs. Tuning the energy gap and edge states of GQDs pave the way toward enormous applications that are not possible by bulk graphene [13- 15].

The electronic and optical properties can be tuned by chemical functionalization of the GQDs. Al-Aqtash and Vasiliev investigations on the electronic properties and geometrical structure of carboxylated graphene showed significant changes in the GQDs structure after the attachment of the COOH group to the surface [16]. Mandal et al. have illustrated that porphyrin functionalized GQDs can be a potential candidate for solar cells [17]. Chen et al. have reported that the photoluminescence properties of GQDs can be tuned precisely by attaching chemical functionalities [18]. Moreover, different functional groups attached to the edge of hexagonal GQDs with zigzag termination have been studied by Y. Li et al. [19]. The authors provide a comparison between the effect of different functional groups attached to the graphene flakes on tuning the electronic and optical properties.

On the other hand molecular modeling with different level of theory is an effective tool to study the physical, chemical and biological properties of carbon nano materials. It could be utilized for some of graphene-derivatives such as fullerene [20] to test the functionality of its surface to act as gas sensor [21-22]. Molecular modeling could be also applied to study the effect of surface modification and biological activity for applications as anti-protease inhibitor [23-24]. Recently, the different functionality of carbon nano materials are reviewed [25]. It is noticed that substitution could affect the biological behavior [26] and electronic properties of fullerene [27-28].

It is clear that carbon nano materials could be directed toward certain application with substitution and/or functionalization with certain functional group. Based upon the above considerations, the present work is conducted to investigate the structure stability and electronic properties of edge carboxylated GQDs. To the best of our knowledge no work has been done on edge functionalization of graphene quantum dots of hexagonal and triangular shapes with armchair and zigzag terminations. We study the effect of attaching different carboxyl groups to the edges of the GQDs on electronic density of states, energy gap, total dipole moment, and IR spectrum.

## 2- LATTICE STRUTURE:

The lattice structure of graphene is made of carbon atoms arranged in hexagonal shape as shown in fig. 1. The lattice vectors can be expressed as: $a_1 = \frac{a}{2}(3,\sqrt{3})$ and $a_2 = \frac{a}{2}(3,-\sqrt{3})$, with $a = 1.42\ A^o$ is the carbon-carbon (C-C) bond length. The arbitrary lattice-translation vector is defined by two whole numbers (n, m) as: $L = na_1 + ma_2$

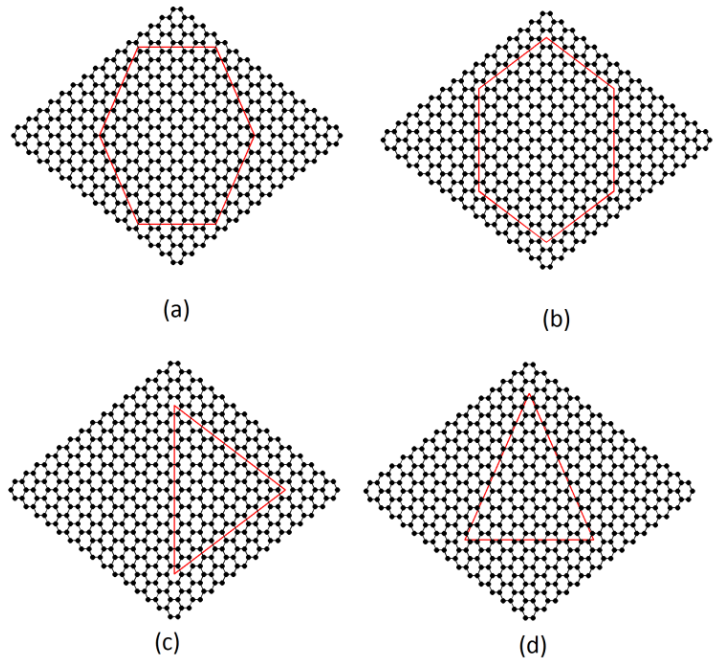

Figure (1): Cutting graphene sheet into small clusters, (a) AHEX, (b) ZHEX,(c) ZTRI and (d) ATRI.

The lattice translation vectors are then used to create the graphene sheet from which we will cut the desired clusters, such as hexagonal and triangular. According to the direction of cutting we can define the edge termination of the clusters, hexagonal and triangular structure with armchair (fig. 1 a, d) and zigzag (fig. 1 b, c) terminations. The cutting of graphene sheet does not always produce structures with perfect armchair or zigzag terminations and some atoms on the edges with single bond to the structure can exist. These atoms are removed from the structure to obtain the desired edge termination. The edges of all the clusters are passivated with hydrogen atoms. Throughout the paper the triangular graphene quantum dot with armchair termination is abbreviated to "ATRI", the triangular with zigzag termination "ZTRI", the hexagonal with armchair termination to "AHEX", the hexagonal with zigzag termination to "ZHEX", and when discussing in general all quantum dots we use GQDs.

## 3- THE COMPUTATIONAL METHOD:

First principles calculations based on DFT [29, 30] have been performed to study electronic properties of the selected GQDs. The DFT is used as implemented in Gaussian 09 [31] that utilizes a basis set of Gaussian type orbital functions. The Becke-three-parameters-Lee-Yang-Parr hybrid functional (B3LYP) [32] is employed in our calculations since it provides a good representation of the electronic structure of graphene-derivatives [33]. The GQDs were fully optimized using the basis set 3-21G [34]. DFT calculations were performed on pyrene ($C_{16}H_{10}$) using the basis sets 3-21G and 6-311G$^{**}$ to justify the sufficiency of the 3-21G basis set for both result accuracy and computational efficiency. It was found that the energy gap using 3-21G and 6-311G$^{**}$ basis sets of pyrene is 3.93 and 3.85 eV respectively. The calculations time is 10 min and 1.36 h respectively. Therefore we consider the basis set 3-21G as an acceptable set for calculations when considering both results accuracy and computation efficiency [35, 36].

## 4- RESULTS AND DISCUSSION:

Modifying the model clusters of graphene is carried out by attaching various carboxyl groups to the edges. It was stated earlier that carboxyl group is one of the most important functional groups in chemistry and biology according to its high dipole moment. The existence of COOH group enhances the reactivity of a given structure and dedicates it to many applications [37]. It could enhance the different properties of a given structure. The following parts will present a study to the effect of COOH group on the stability and electronic properties of the selected clusters.

## 4.1- STABILITY:

The optimized structures of ATRI, ZTRI, AHEX , and ZHEX before and after attachment of carboxyl groups to the edges are presented in figures 2, 3. The optimization was done by minimizing the total energy at the B3LYP/3-21G level of calculation without symmetry constraint . The calculated C-C bond lengths of all the selected clusters before adding carboxyl group are in the range from 1.36 to 1.46 ($A^0$) which is comparable to the C-C bond length (1.42 $A^0$) in the bulk graphene. The values of bond lengths which are slightly differ from 1.42 ($A^0$) such as 1.36 represent the bond length between atoms at the edge while 1.46 represents bond length between bulk atoms. The attachment of COOH groups to the corners of GQDs have almost a negligible effect on the C-C bond lengths

(see table 1) of the original structure which means that the attachment of COOH to the corners of hexagonal and triangular graphene does not distort the structure. The optimal C-COOH distances ($d_{cx}$) for different clusters are also presented in table1. The distances between the attached COOH and different GQDs are comparable to each other with lowest distance 1.475 $A^0$ in ATRI and highest 1.484 $A^0$ in ZHEX.

All the optimized structures were used to calculate the frequiecies. It is found that for all the clusters the frequiencies are real (positive) which confirm the stability of the selected systems (see sections 4).

Table (1): The C-C bond lengths and the binding energies of the selected GQDs functionalized with different COOH groups to the edges.

| Structure | | $d_{cx}$ ($A^0$) | $d_{cc}$ ($A^0$) | $E_b$ (eV) |
|---|---|---|---|---|
| ATRI ($C_{60}H_{24}$) | COOH | 1.475 | 1.372-1.459 | 0.157 |
| | 2 COOH | 1.475, 1.476 | 1.372-1.459 | 0.304 |
| | 3 COOH | 1.476 | 1.372-1.459 | 0.441 |
| ZTRI ($C_{46}H_{18}$) | COOH | 1.480 | 1.392-1.443 | 0.204 |
| | 2 COOH | 1.480 | 1.375-1.444 | 0.391 |
| | 3 COOH | 1.499, 1.498 | 1.393-1.461 | 0.563 |
| AHEX ($C_{42}H_{18}$) | COOH | 1.477 | 1.387-1.464 | 0.505 |
| | 2 COOH | 1.477 | 1.387-1.464 | 0.965 |
| | 3 COOH | 1.478, 1.4781 | 1.387-1.464 | 1.385 |
| | 4 COOH | 1.478 | 1.387-1.464 | 1.769 |
| | 5 COOH | 1.478, 1.4781, 1.479,1.4791 | 1.388-1.463 | 2.123 |
| | 6 COOH | 1.479 | 1.389-1.463 | 2.450 |
| ZHEX ($C_{54}H_{18}$) | COOH | 1.481 | 1.362-1.455 | 0.182 |
| | 2 COOH | 1.482 | 1.362-1.455 | 0.351 |
| | 3 COOH | 1.482, 1.483 | 1.361-1.456 | 0.508 |
| | 4 COOH | 1.482, 1.483,1.484 | 1.344-1.455 | 0.654 |
| | 5 COOH | 1.482, 1.483, 1.484 | 1.361-1.456 | 0.7898 |
| | 6COOH | 1.483, 1.484 | 1.369-1.456 | 0.9161 |

In addition, the binding energy of GDQs was calculated to compare the stabilities of different shapes. $E_b$ is calculated as: $E_b = (N_C E_C + N_H E_H - E_{C-H})/N$. With $N_C$, $N_H$, and $N$ are the numbers of carbon, hydrogen, and carbon plus hydrogen atoms respectively. $E_C$ and $E_H$ are the ground state energies of the isolated carbon and hydrogen atoms (calculated using the same method). $E_{C-H}$ is the ground state energy of the GQDs.

The binding energies per atom for AHEX ($C_{42}H_{18}$) and ZTRI ($C_{46}H_{18}$) are almost the same ($E_b$=7.05 (eV) for AHEX and $E_b$= 7.12 (eV) for ZTRI) because they have similar numbers of carbon and hydrogen atoms [39]. However, for the structure having almost the same number of C-atoms but different number of H- atoms, such as ZHEX ($C_{54}H_{18}$) and ATRI ($C_{60}H_{24}$), the calculated binding energies are 7.4 and 7.2 eV respectively. The ZHEX require lower number of hydrogen atoms ($H_{18}$) than ATRI ($H_{24}$) for passivation. The lower number of H- atoms in ZHEX is the reason for increasing the binding energy.
The binding energy of COOH groups on GQDs is calculated as the difference between the ground state energy of the carboxylated GQDs and the sum of the ground state energies of GQDs and COOH groups. All the calculated $E_b$, as in table 1, are positive which insure the stability of the carboxylated structures. Also, the binding energy significantly increases by increasing the number of attached COOH to the GQDs edges. The highest stability is observed in carboxylated AHEX, namely $E_b$=2.5 (eV) for AHEX + 6 COOH.

The total dipole moment was claculated for all the GQDs flakes. It is observed (see table2) that the net dipole moment for non functionalized ZTRI and ATRI has a finite value while for ZHEX and AHEX it has zero value. This can be explained in terms of the fact that the carbon atom will absorb part of the electron of the H atom making it positively charged with respect to the negatively charged carbon atoms. Thereofore, the GQDs will have permanent polarization and the total dipole moment will be determined from the summation over all local dipoles at the edges. Consequantly, the total dipole moment will depend on the geometrical shape of the flake.

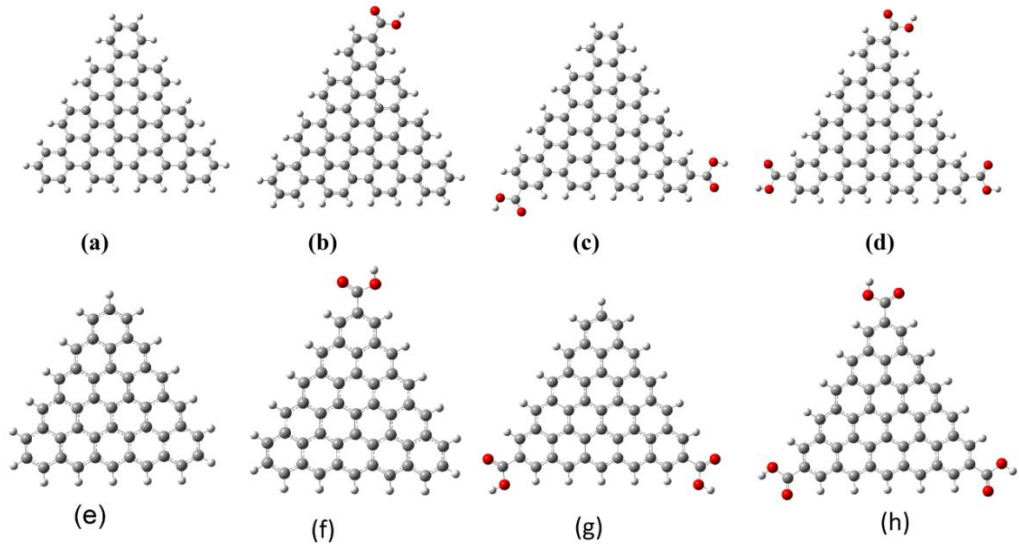

Figure (2): The optimized structures of ATRI (a, b, c, d) and the optimized structure of ZTRI (e, f, g, h) without and with edge functionalization with carboxyel groups.

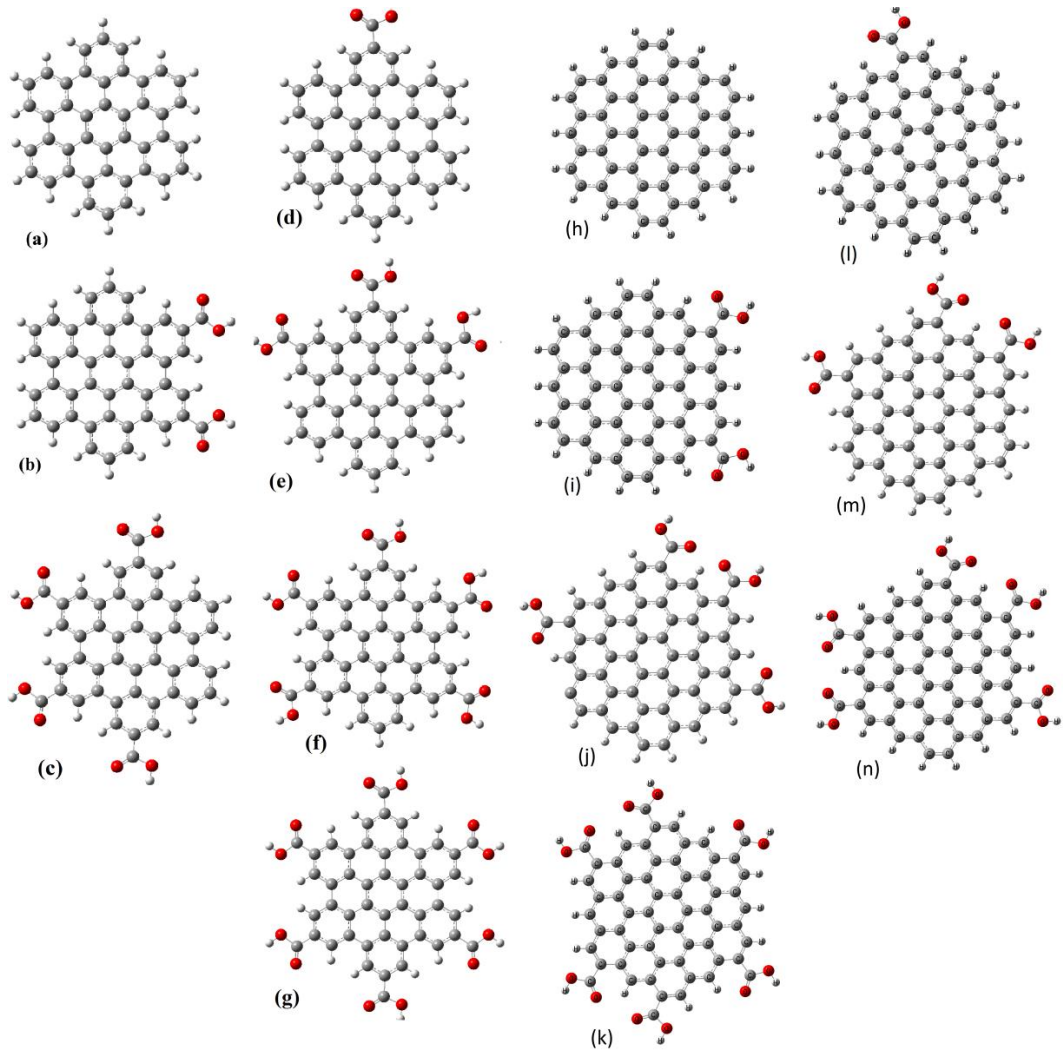

Figure (3): (a - g) Relaxed structures of AHEX and (h - n) ZHEX optimized structures, without and with edge carboxylation.

Attachment of COOH groups to different GQDs significantly increases the total dipole moment (TDM), see table 2. Triangular structures have the highest TDM ~ 4 (Debye) because they have an original none zero TDM. The increase of the dipole moment with increasing the number of COOH is not systematic, i.e. for ATRI (ZTRI) the highest value of TDM obtained by attaching 2 COOH (COOH) to its corners.

Table 2. The structure of graphene clusters with the corresponding energy gap (ΔE) and total dipole moment (TDM).

| | Structure | ΔE (eV) | TDM |
|---|---|---|---|
| ATRI ($C_{60}H_{24}$) | Without COOH | 3.269 | 0.001 |
| | COOH | 3.195 | 2.701 |
| | 2 COOH | 3.182 | 4.033 |
| | 3 COOH | 3.136 | 3.81 |
| ZTRI ($C_{46}H_{18}$) | Without COOH | 0.271 | 0.0035 |
| | COOH | 0.311 | 3.897 |
| | 2 COOH | 0.307 | 0.110 |
| | 3 COOH | 0.269 | 2.32 |
| AHEX ($C_{42}H_{18}$) | Without COOH | 3.668 | 0.000 |
| | COOH | 3.542 | 2.021 |
| | 2 COOH | 3.496 | 1.587 |
| | 3 COOH | 3.520 | 3.037 |
| | 4 COOH | 3.459 | 3.712 |
| | 5 COOH | 3.435 | 3.414 |
| | 6 COOH | 3.447 | 2.328 |
| ZHEX ($C_{54}H_{18}$) | Without COOH | 2.886 | 0.000 |
| | COOH | 2.762 | 1.855 |
| | 2 COOH | 2.742 | 0.011 |
| | 3 COOH | 2.718 | 1.595 |
| | 4 COOH | 2.69 | 2.062 |
| | 5 COOH | 2.638 | 3.374 |
| | 6COOH | 2.610 | 2.993 |

## 4.2- DENSITY OF STATES AND HOMO-LUMO ENERGY GAP:

Density of states (DOS) has been calculated for ATRI, ZTRI, AHEX, and ZHEX before and after edge-functionalization where each electronic state was represented in terms of Gaussian function $\frac{1}{\sqrt{2\pi}\alpha}\exp\left[\frac{(\varepsilon-\varepsilon_i)^2}{2\alpha^2}\right]$ with broadening $\alpha = 0.02 \ eV$. The Fermi energy is seated at zero by defining $E_F = (E_{HOMO}+E_{LUMO})/2$. Figure 4 (a) shows the DOS of ATRI functionalized with COOH at the corners of the triangle. The effect of attaching COOH to the ATRI flake is the appearance of new peaks in the spectrum. The peak shown in fig. 4a (under the arrow) is a result of the addition of COOH groups and its intensity increases by increasing the number of carboxyl group. The HOMO-LUMO energy gap slightly decreases by the addition of COOH groups, $E_g$ decreases from 3.3 to 3.1 eV as given in table 2. Figure 4 (b) presents the DOS of the carboxylated ZTRI, in contrast to ATRI, the energy gap is very small.

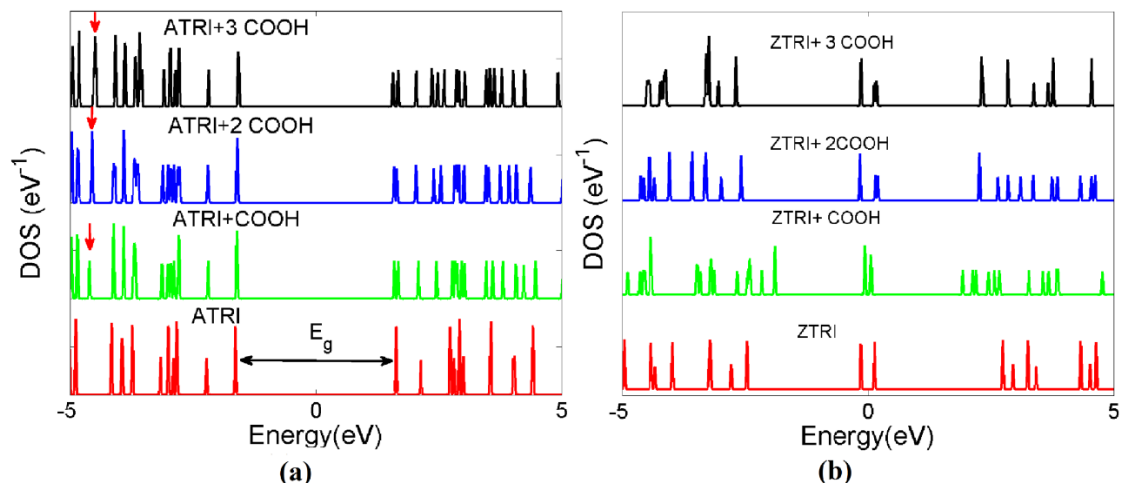

Figure (4): Density of states of ATRI cluster (a) and ZTRI (b) without and with the addirion of COOH groups in the edges of the triangule.

Figure 5 shows the HOMO and LUMO for ATRI (fig. 5 a) and ZTRI (fig.5 b) graphene flakes. It is observed that for ATRI the orbitals distribute over the surface and edge, providing delocalized HOMO and LUMO states. The effect of addition of carboxyl group to the ATRI is only the redistribution of the HOMO and the LUMO over the structure surface and edges. This distribution illustrates that the HOMO and LUMO in ATRI are not edge states. The edge effect can be found in ZTRI, where HOMO is distributed over the edges and LUMO is distributed over the center of the flake. Therefore, these edge states in ZTRI that fill the low energy region are the reason for the small energy gap in triangular graphene with zigzag termination. The density of states AHEX and ZHEX before and after adding carboxyl groups are shown in figure (6). Similar to triangular structures, new peaks in the DOS spectrum emerge due to the addition of COOH groups where there number and intensities depends on the number of the attached groups. A fluctuation in the energy gap ($E_g$) is observed as a function of the number of attached carboxyl group with overall slight decrease in $E_g$ (in the range from Eg=3.7 to 3.4 for AHEX and $E_g$=2.9 to 2.6 for ZHEX as seen in table 2). The energy gap strongly depends on the shape and edge termination of the graphene flake, AHEX has a considerable higher energy gap than the ZHEX. Also it is already reported that $E_g$ decreases in GQDs of any shape with increasing its size [3, 38] therefore we didn't discuss the dependence on size here.

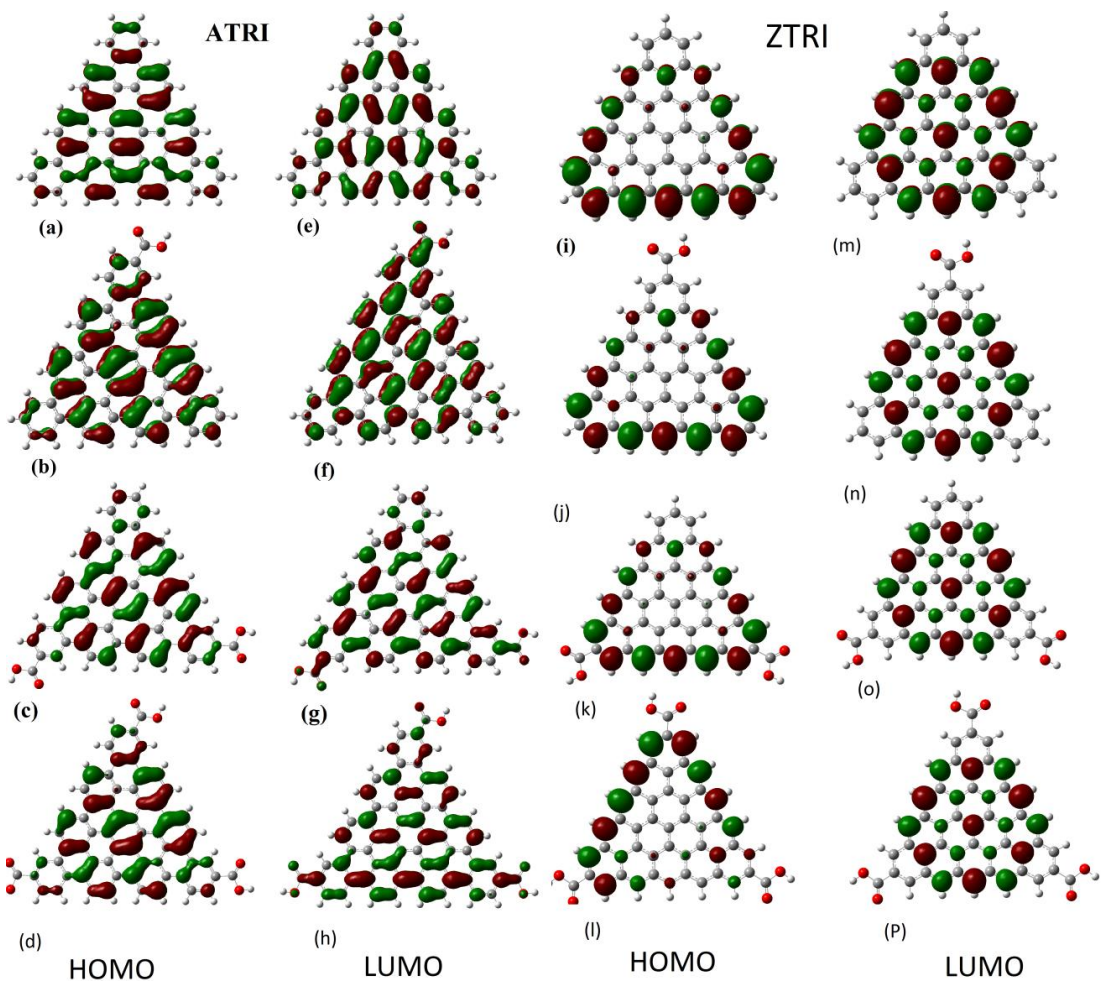

Figure (5): HOMO/LUMO distribution calculated for ATRI and ZTRI graphene flakes at the B3LYP/3-21G level.

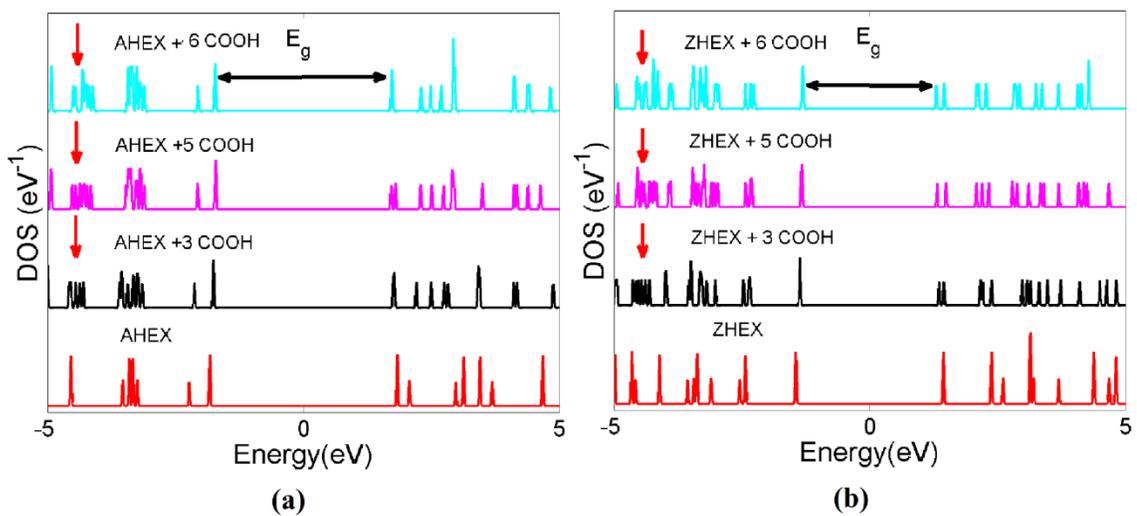

Figure (6): DOS of AHEX (a) and ZHEX (b) before and after carboxylation.

THE HOMO/LUMO for AHEX and ZHEX are shown in figure (7). The calculations are made only for AHEX or ZHEX without attachment of COOH and with attachement of three, five, and six groups. Again as observed in ATRI, the HOMO/LUMO are distributed over the surface and the edges of the structure. Therefore for ATRI, AHEX, ZHEX the HOMO and LUMO are bulk delocalized states that are distributed over all the surface and edge of the flake. While for ZTRI , the HOMO is edge localized state that exist on the edges and the LUMO distribute over the center of the flake.

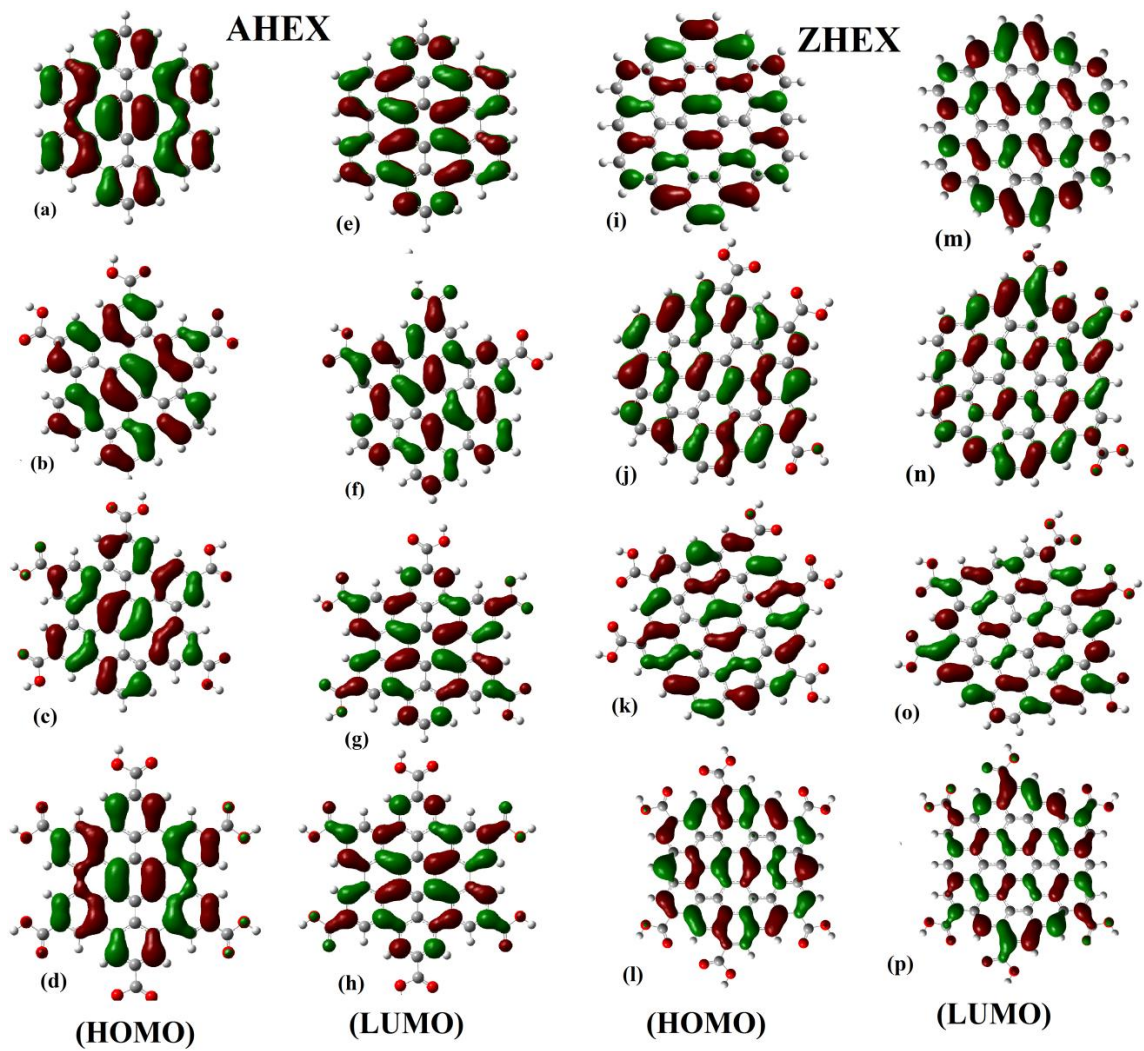

Figure (7): HOMO/LUMO distribution for AHEX and ZHEX calculated at the B3LYB/3-21G level.

## 4.3- INFRARED SPECTRA

The B3LYB/3-21G caculated IR spectra are indicated in Figures 8, 9, 10 , and 11 for ATRI, ZTRI, AHEX, and ZHEX respectrively before and after attaching carboxyl groups to the edges. The aim of employing the vibrational sepectra is to confirm that the above caculated electronic properites are carried out upon optimized strcutures. Hence, second dervative calculations in this work confirm all the data obtained through first dervative. The figures indicate that there are no negative frequencies, accordinlgy the data given in the above sections are confirmed with the obtained positive frequencies.

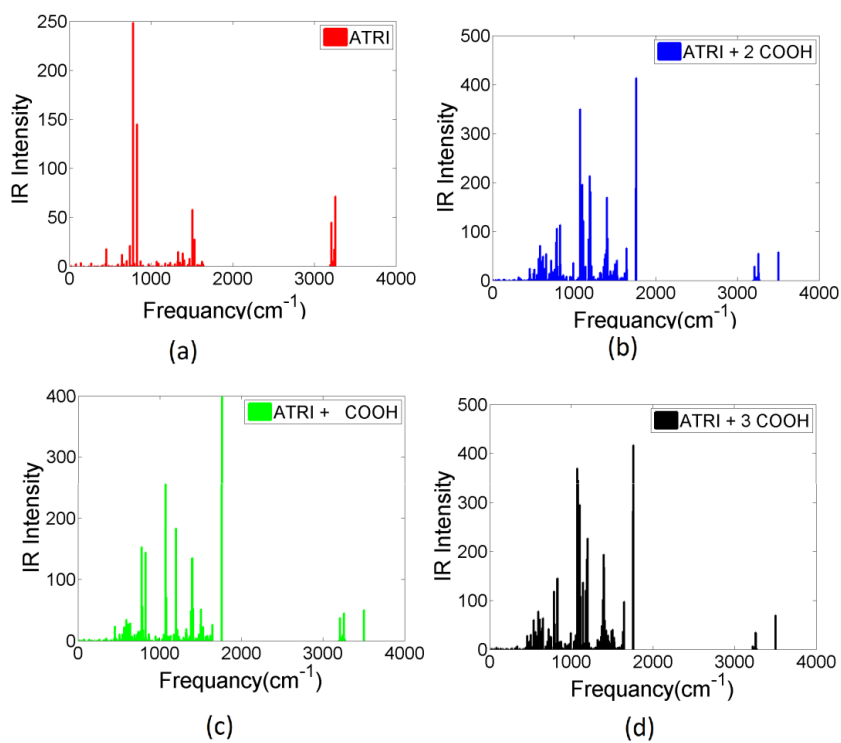

Figure (8): IR spectra of ATRI without COOH (a) and with COOH groups attached to the edges (b, c, d).

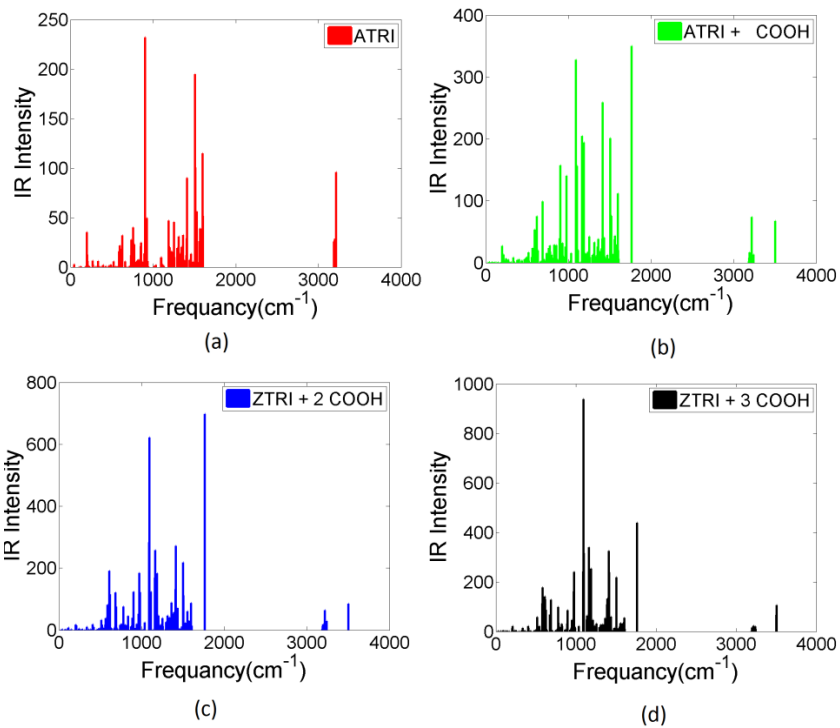

Figure (9): The IR spectra for ZTRI without and with carboxylation.

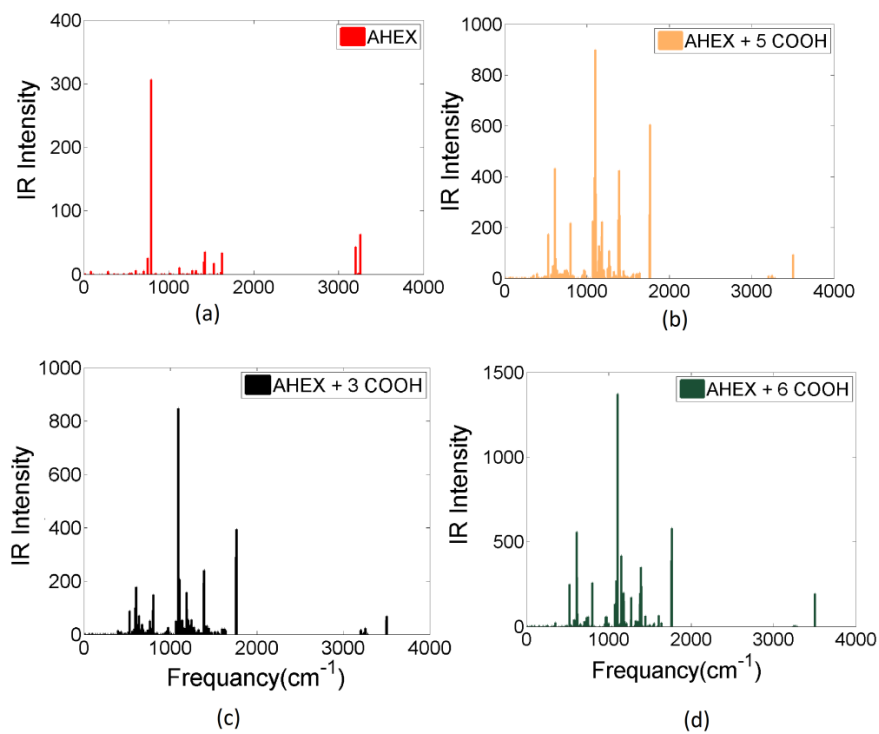

Figure (10): IR intensity of AHEX before and after attachemnet of COOH groups.

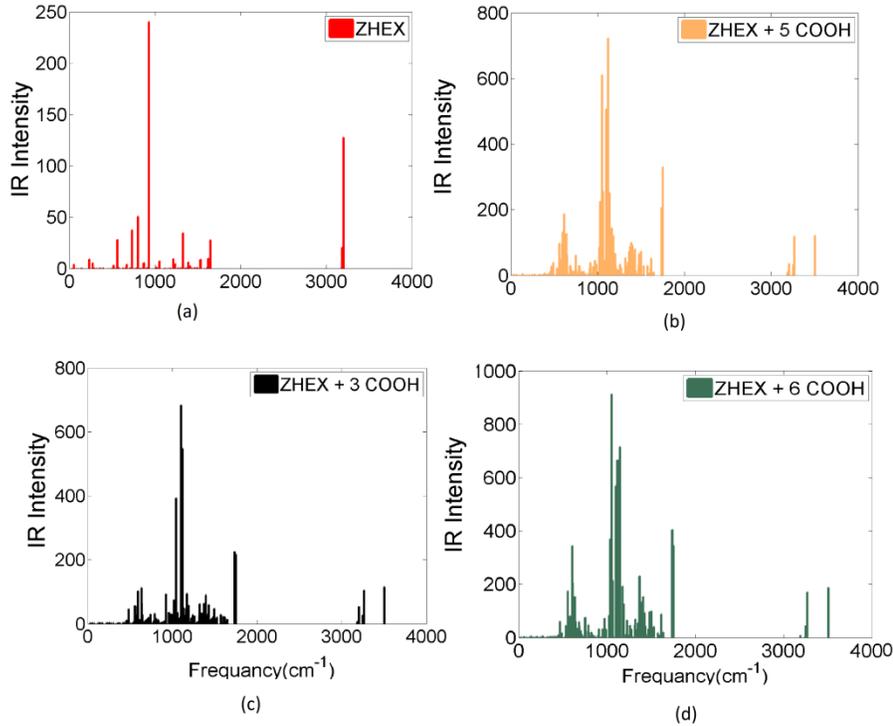

Figure (11): The IR spectra for ZHEX before and after attaching COOH groups.

## CONCLUSION:

Using density functional theory the structure stability and electronic properties were investigated for graphene quantum dots functionalized with carboxyl groups. The calculations are conducted on hexagonal and triangular GQDs with armchair and zigzag terminations. The positive values of the obtained binding energies for all the selected flakes confirm their stability. The total dipole moments for hexagonal clusters are zero while for triangular ones it has finite values. This is shape effect where for the hexagonal shapes the local dipoles moment at the six edges cancel each other, while for the triangular clusters there will be a net dipole moment coming from one of the local dipoles at the three edges.

The stability of the carboxylated clusters increases by increasing the number of carboxyl groups. The attachment of COOH significantly increases the total dipole moment with very high values observed in triangular structures, the ATRI (4.033 Debye). The edge states appear in triangular GQDs with zigzag edges are the reason for the completely different energy spectrum from that in ATRI, AHEX, and ZHEX. HOMO in ZTRI is localized in the edges which is different from other clusters where their HOMO orbitals are delocalized and distributed over the surface, also, the ZTRI energy gap is very small compared to other clusters. The HOMO/LUMO band gap slightly changed by carboxylation with no trend even we could say it almost remains unchanged in some structures. The increase in the total dipole moment reflects the fact that the given structures are highly reactive with their surrounding media. It could be concluded that, carboxyl group significantly enhances the studied properties of GQDs by increasing their reactivity which dedicate its surface in many applications such as sensors.

## ACKNOWLEDGEMENT:

The authors are very grateful to Prof. A. Fakhry for a careful reading of the manuscript and the valuable corrections.